\documentstyle[12pt]{article}
\begin{document}

\begin{center}
\vglue 1cm
{\Large\bf Leptophobic $Z'$ in stringy flipped SU(5)\\}
\vglue 1.5cm
{\Large Jorge L. Lopez$^1$ and D.V. Nanopoulos$^{2,3}$}
\vglue 1cm
\begin{flushleft}
$^1$Department of Physics, Bonner Nuclear Lab, Rice University\\ 6100 Main
Street, Houston, TX 77005, USA\\
$^2$Center for Theoretical Physics, Department of Physics, Texas A\&M
University\\ College Station, TX 77843--4242, USA\\
$^3$Astroparticle Physics Group, Houston Advanced Research Center (HARC)\\
The Mitchell Campus, The Woodlands, TX 77381, USA\\
\end{flushleft}
\end{center}

\vglue 1.5cm
\begin{abstract}
We show that leptophobic $Z'$ gauge bosons occur naturally in flipped SU(5) and
may shift $R_b$ in an interesting way without upsetting the good values of
$\Gamma_{\rm had}$ and $R_c$. Within a string-derived version of the model, we study three possible scenarios and the constraints imposed on model building that would allow the new symmetry to remain unbroken down to low energies. Such $Z'$ gauge boson has generation non-universal couplings to quarks that violate
parity maximally in the up-quark sector, and may contribute significantly to
spin asymmetries in polarized $pp$ scattering experiments now being prepared
for RHIC.
\end{abstract}
\vspace{1cm}
\begin{flushleft}
E-mail addresses:\\
\small\tt
\baselineskip=12pt
lopez@physics.rice.edu\\
dimitri@phys.tamu.edu\\
\end{flushleft}
\newpage

\setcounter{page}{1}
\pagestyle{plain}
\baselineskip=14pt

\section{Introduction}
\label{sec:introduction}
Recent experimental results concerning the rate of $b$ quark production
at LEP~1 ($R^{\rm exp}_b=0.2179\pm0.0012$ \cite{LEP}, $R^{\rm SM}_b=0.2157$) and the distribution of high-$E_T$ jets at the Tevatron (as reported by the CDF Collaboration \cite{CDF}), that may indicate the first departure from Standard Model expectations, have revived interest in $Z'$ models as possible beyond-the-Standard-Model explanations for these phenomena \cite{Altarelli,CL,others,Babu,Stirling,Frampton,Barger,Alon}.
In order not to disturb the agreement with Standard Model expectations in the
lepton sector at LEP, the new $Z'$ scenarios call for a leptophobic $Z'$ that
mixes with the regular $Z$.\footnote{If the $Z'$ is also to explain the CDF
data, it should be unusually broad and heavy \cite{Altarelli}. As D0 has
studied a much larger data sample than that reported by CDF, and no anomalous
distributions have been observed \cite{D0}, we do not consider $Z'$ gauge
bosons with such characteristics here. For a comparison of the CDF and D0
results see Ref.~\cite{Tung}.} Such a $Z'$ may shift $R_b$ in an interesting
way if its couplings to quarks are suitably chosen. However, one must insure that the total hadronic width ($\Gamma_{\rm had}$) remains essentially unchanged relative to Standard Model expectations, as it agrees rather well with observations. A new constraint in this class of models has recently arisen, in that the $R_c$ ratio is now found to agree rather well with Standard Model expectations ($R^{\rm exp}_c=0.1715\pm0.0012$ \cite{LEP}, $R^{\rm SM}_c=0.172$). Traditional $Z'$ searches at hadron colliders ({\em i.e.}, via its decay into charged leptons \cite{Z'searches}) are not sensitive to a leptophobic $Z'$, whereas $Z'\to jj$ searches \cite{UA2,Harris} are hampered by much larger backgrounds, although excluded regions of parameter space can still be obtained. The influence of a leptophobic $Z'$ may also be felt via additional contributions to the top-quark cross section \cite{Stirling}, enhanced $b\bar b$ event yields at the Tevatron \cite{Barger}, and spin asymmetries in polarized $pp$ collisions at RHIC \cite{TV}.

If such a $Z'$ explanation to the seemingly anomalous data is to be taken
seriously, one must provide a consistent theoretical framework where the new
gauge boson and its required properties arise naturally. In our minds that
should be taken to be string model-building, where the thorny question of
cancellation of anomalies is dealt with automatically. New light neutral gauge
bosons were early on considered to be the ``smoking guns" of string, back when
$E_6$ was the proverbial string-inspired gauge group \cite{HR}. The popularity
of string $Z'$s however waned as non-grand-unified gauge groups became the
natural outcome of actual string-derived models, such as SU(5)$\times$U(1)
[``flipped SU(5)"], SU(3)$^3$, SU(4)$\times$SU(2)$\times$SU(2) [``Pati-Salam"
models], and SU(3)$\times$SU(2)$\times$U(1) [``Standard-like" models]. In these
models one in fact has an excess of U(1) gauge groups at the string scale,
but they typically get broken in the vacuum shifting process required to
cancel the anomalous $\rm U_A(1)$ that is characteristic of this type of string
constructions. It is then interesting to explore whether string models
can accomodate such gauge bosons at all, a search that is likely to lead to new
and restrictive constraints on string model building. In this paper we explore
this question in the context of stringy flipped SU(5).

We first show that leptophobia is very natural in flipped SU(5), and that it
may provide the shift in $R_b$ that seems to be preferred experimentally
(Sec.~\ref{sec:RbRc}), while at the same time keeping $\Gamma_{\rm had}$ and
$R_c$ essentially unchanged. We then explore three scenarios for possible leptophobic $Z'$ gauge bosons in a string-derived version of flipped SU(5)
(Sec.~\ref{sec:scenarios}). Our second scenario is particularly compelling,
entailing generation non-universal $Z'$ couplings to quarks; it is further
studied in Sec.~\ref{sec:revisited}. We also discuss the current experimental
limits on and the prospects for detecting such $Z'$ gauge bosons, which possess
parity-violating $Z'$ couplings to up-type quarks, and may yield observable
contributions to spin asymmetries in polarized $pp$ scattering experiments
being prepared for RHIC (Sec.~\ref{sec:exp}). Sec.~\ref{sec:conclusions}
summarizes our conclusions.

\section{Flipped leptophobic $Z'$ and $R_b$}
\label{sec:RbRc}
Let us first point out that leptophobia is very {\em natural} in
SU(5)$\times$U(1), where the particle content of the Standard Model is
contained in the representations
\begin{equation}
F=({\bf10},{\textstyle{1\over2}})=\{Q,d^c,\nu^c\},\quad
\bar f =(\bar{\bf5},-{\textstyle{3\over2}})=\{L,u^c\},\quad
\ell^c=({\bf1},{\textstyle{5\over2}})=\{e^c\}.
\label{eq:reps}
\end{equation}
The new $Z'$ would be leptophobic if $\bar f$ and $\ell^c$ (which contain the
Standard Model leptons) are uncharged under the new U(1), while most of the
quarks (in $F$) could still couple to it. In contrast, in regular SU(5)
($F=\{Q,u^c,e^c\}$, $\bar f=\{L,d^c\}$), SO(10), or $E_6$ such a separation
(symmetry-based leptophobia) is not possible. However, leptophobia may still be
achieved dynamically under certain circumstances, as demonstrated in
Ref.~\cite{Babu} in the case of the subgroup of $E_6$ called the
``$\eta$-model" \cite{EENZ}.

The effect of a $Z'$ mixing with the regular $Z$ and its impact on LEP physics
has been addressed previously, particularly in the context of $R_b,R_c$
\cite{Altarelli,others,Babu,Barger}. Here we present a succinct discussion,
emphasizing the flipped SU(5) novelties. Small $Z$-$Z'$ mixing amounts to a
shift in the vector and axial-vector couplings of the standard $Z$:
\begin{equation}
C_V=C_V^0+\theta(g_{Z'}/g_Z)\,C'_V\,;\qquad
C_A=C_A^0+\theta(g_{Z'}/g_Z)\, C'_A\ ,
\label{eq:shifts}
\end{equation}
where $\theta$ is the small $Z$-$Z'$ mixing angle, $g_Z=g/\cos\theta_w$ is the
usual weak coupling, $g_{Z'}$ is the new $\rm U'$ gauge coupling, $C^0_{V,A}$
are the usual vector and axial-vector $Z$ couplings to fermions, and $C'_{V,A}$
are related to the charges of the fermions under $\rm U'$. As
Eq.~(\ref{eq:reps}) shows, a leptophobic $Z'$ in flipped SU(5) violates parity
in the up-quark sector, but not in the down-quark sector. Using
$C'_V=Q(\psi_L)+Q(\psi_R)$ and $C'_A=-Q(\psi_L)+Q(\psi_R)$ we obtain
\begin{equation}
\begin{tabular}{c|crcccc}
&$C^0_V$&$C^0_A$&$Q(\psi_L)$&$Q(\psi_R)$&$C'_V$&$C'_A$\\ \hline
up&${1\over2}-{4\over3}x_w$&${1\over2}$&$c$&0&$c$&$-c$\\
down&$-{1\over2}+{2\over3}x_w$&$-{1\over2}$&$c$&$c$&$2c$&0\\
\end{tabular}
\label{eq:charges}
\end{equation}
where $x_w=\sin^2\theta_W$, and $c$ is the $\rm U'$ charge of the
$F=({\bf10},{1\over2})$ representation. These results apply to each individual
generation, which generally could have different charges under $\rm U'$, {\em i.e.}, $c_{1,2,3}$.

With the above information, which depends solely on the flipped SU(5) origin
of the leptophobic $Z'$, one can obtain the first-order shifts in
$\Gamma_{b\bar b}=\Gamma(Z\to b\bar b)$, $\Gamma_{c\bar c}=\Gamma(Z\to c\bar
c)$ and
$\Gamma_{\rm had}=\Gamma(Z\to{\rm hadrons})=\Gamma_{d\bar d}+\Gamma_{s\bar
s}+\Gamma_{b\bar b}+\Gamma_{u\bar u}+\Gamma_{c\bar c}$ (and therefore
$R_b,R_c$) in terms of the composite parameter
\begin{equation}
\delta\equiv\theta(g_{Z'}/g_Z)\ .
\label{eq:delta}
\end{equation}
Recalling that $\Gamma(Z\to q\bar q)=\Gamma_0(C^2_{V_q}+C^2_{A_q})$, where
$\Gamma_0=G_F M^3_Z/(2\sqrt{2}\pi)$, one can easily determine the relevant
first-order shifts ({\em i.e.}, linear in $\delta$) for arbitrary choices of the $c_{1,2,3}$ charges.

What should the $\rm U'$ charges of the SU(5) multiplets be? This question
can be answered exactly in explicit string models, as we discuss in
Sec.~\ref{sec:second}. For now let us just state our choices:
\begin{equation}
\begin{tabular}{lrclrclr}
$F_0$&$-{\textstyle{1\over2}}$&\qquad&$\bar F_4$&${1\over2}$&\qquad&$\bar f_{2,3,5}$&0\\
$F_1$&$-{\textstyle{1\over2}}$&&$\bar F_5$&0&&$\ell^c_{2,3,5}$&0\\
$F_2$&0\\
$F_3$&1\\
$F_4$&$-{\textstyle{1\over2}}$
\end{tabular}
\label{eq:Fcharges}
\end{equation}
These charge choices reflect some principles: the leptons (in $\bar f_{2,3,5},
\ell^c_{2,3,5}$) are uncharged (leptophobia); there is a pair of
$({\bf10},\overline{\bf10})$ ($F_2,\bar F_5$) whose neutral components acquire GUT scale vevs and break SU(5)$\times$U(1) in the standard way, and by virtue of being uncharged leave $\rm U'$ unbroken; the remaining
representations enforce $\rm Tr\,U'=0$ and contain three generations of quarks
and an extra $({\bf10},\overline{\bf10})$ to allow string unification.
The $c_{1,2,3}$ are then to be taken from the set
$\{0(1),1(1),-{1\over2}(3)\}$, where the number in parenthesis
indicates the number of generations that may carry such charge.

\begin{figure}[p]
\vspace{6in}
\includegraphics{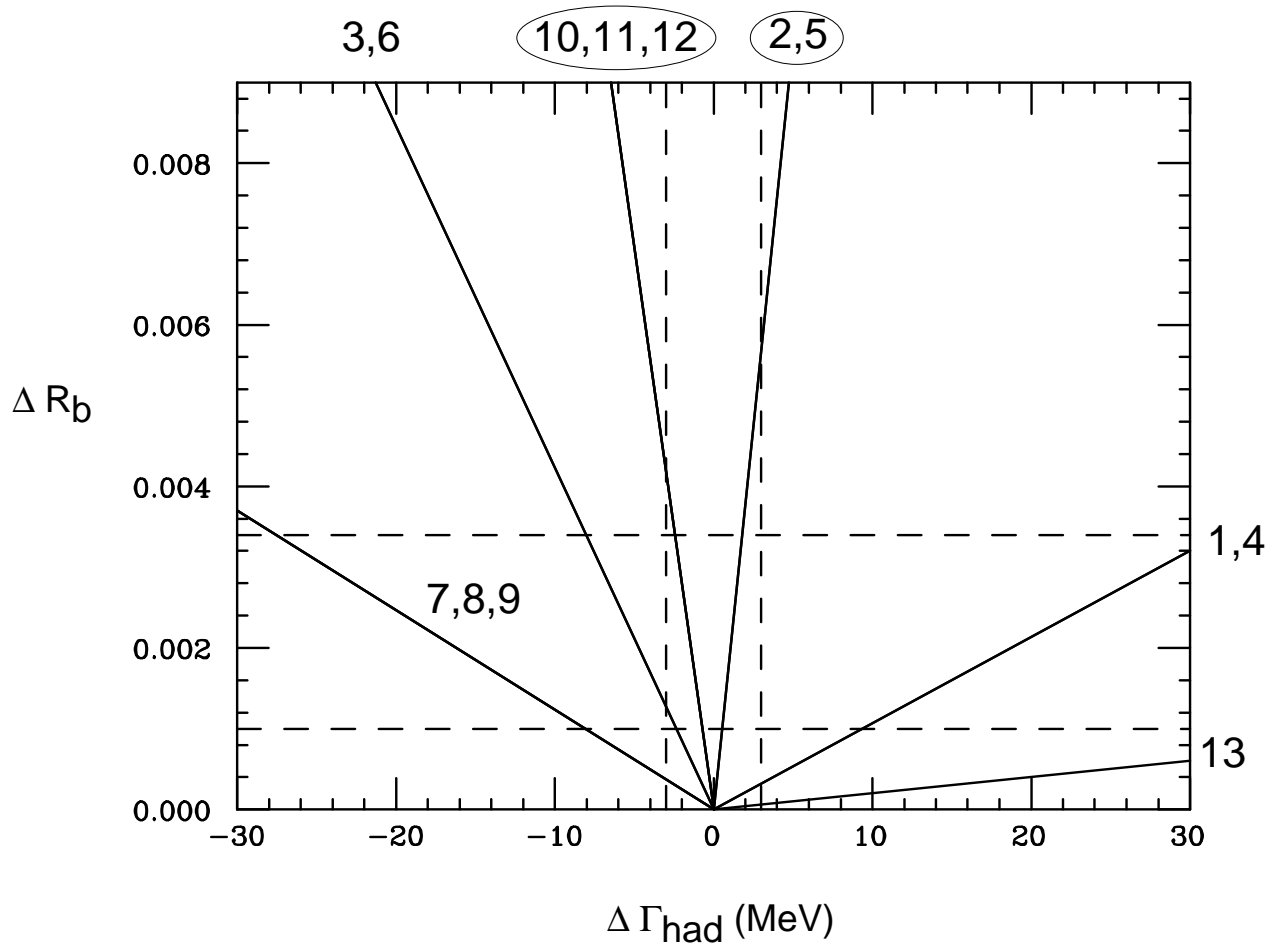}
\caption{The correlated shifts in $R_b$ and $\Gamma_{\rm had}$ in flipped SU(5)
for the various $\rm U'$ charge assignment combinations shown in Table~1. The
dashed lines delimit the experimental uncertainty in $\Gamma_{\rm had}$ and
the required shift in $R_b$ to fall within the experimental limits. The circled
charge assignments (2,5,10,11,12) agree with experiment.}
\label{fig:RbGhad}
\end{figure}

In Table~\ref{Table0} we display the 13 possible charge combinations, along
with the fractional changes in $\Gamma_{\rm had}$, $R_b$, and $R_c$ in units
of $\delta$. The value of $\delta$ is constrained by its effect on the
many electroweak observables, where it enters through a tree-level correction
to the $\rho$ parameter due to the $Z$-$Z'$ mixing. Detailed fits to the
electroweak data allow values of $|\delta|$ as large as $\sim10^{-2}$
\cite{Altarelli,Babu,Barger}. For the representative choice $\delta=0.01$, in
Table~\ref{Table0} we also display the actual shifts in $\Gamma_{\rm had}$ (in
MeV), $R_b$, and $R_c$, all of which scale linearly with $\delta$. To help
decide which of the charge assignments satisfy the present experimental
constraints, in Fig.~\ref{fig:RbGhad} we plot the correlated values of $\Delta
R_b$ versus $\Delta\Gamma_{\rm had}$, for each of the 13 charge assignments
(the lines are parametrized by $\delta$). Note that different charge
assignments may yield the same $\Delta R_b$-$\Delta\Gamma_{\rm had}$
correlation. Since the LEP measured value of $\Gamma_{\rm had}$ agrees with the
Standard Model prediction within 1~MeV, and the experimental uncertainty in
$\Gamma_{\rm had}$ is 3~MeV, we constrain possible charge assignments by
requiring $|\Delta\Gamma_{\rm had}|<3\,{\rm MeV}$, as denoted by the vertical
dashed lines in Fig.~\ref{fig:RbGhad}. Furthermore, we demand that $\Delta R_b$
be in the interval $0.0010-0.0034$ (denoted by the horizontal dashed lines in
Fig.~\ref{fig:RbGhad}), such that when added to the Standard Model prediction
($R_b^{\rm SM}=0.2157$, for $m_t=175\,{\rm GeV}$) the resulting $R_b$ falls within the experimental allowed window:\footnote{Such level of precision should suffice, as one still needs to include one-loop supersymmetric corrections to $R_b$, which are not expected to be enhanced \cite{LEP15}, especially in supergravity models \cite{WLN}. In any event, shifting $\Delta R_b$ somewhat does not change the set of allowed charge assignments.} $R^{\rm exp}_b=0.2179\pm0.0012$ \cite{LEP}. From Fig.~\ref{fig:RbGhad} we conclude that only charge assignments 2,5,10,11,12 satisfy the experimental requirements of $R_b$ and $\Gamma_{\rm had}$.

\begin{figure}[p]
\vspace{6in}
\includegraphics{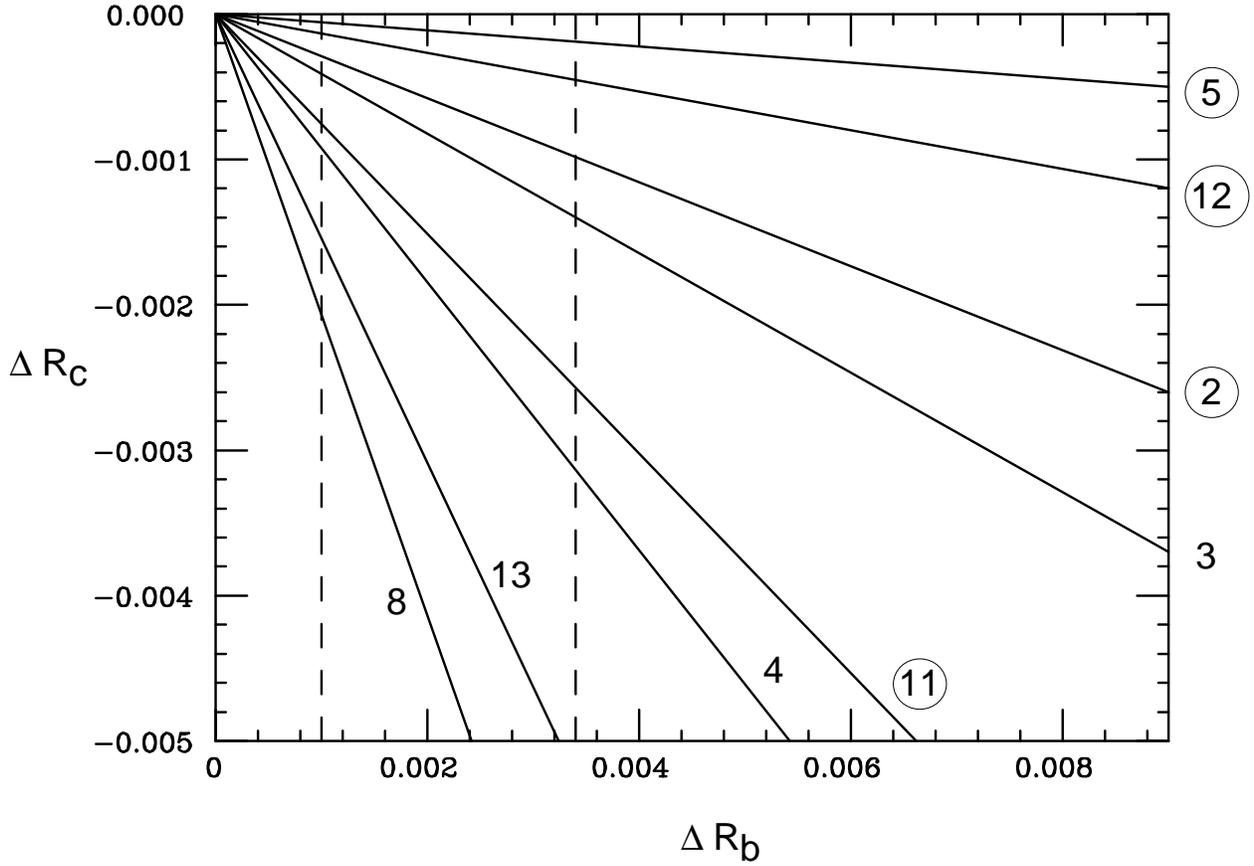}
\caption{The correlated shifts in $R_c$ and $R_b$ in flipped SU(5)
for the various $\rm U'$ charge assignment combinations shown in Table~1.
(Combinations not shown entail shifts in $R_c$ and $R_b$ in the same
direction.) The dashed lines delimit the required shift in $R_b$ to fall within
the experimental limits. The circled charge assignments (2,5,11,12) agree with
experimental data on $\Gamma_{\rm had}$ (see Fig.~1).}
\label{fig:RcRb}
\end{figure}

To consider the effect of $R_c$, in Fig.~\ref{fig:RcRb} we plot the correlated
values of $\Delta R_c$ versus $\Delta R_b$. The latest experimental value of $R_c$ ($R^{\rm exp}_c=0.1715\pm0.0056$ \cite{LEP}) is now in good agreement with the Standard Model prediction ($R^{\rm SM}_c=0.172$).  Only cases 2,5,11,12 satisfy the experimental requirements from $R_b$, $R_c$, and $\Gamma_{\rm had}$.\footnote{As the data shows preference for $\Delta R_c<0$ and $\Delta R_b>0$, we exclude from further consideration charge assignments that entail shifts in the same direction, {\em i.e.}, cases 1,6,7,9,10.}
It is important to note that these charge assignments predict modest shifts in $R_c$: $|\Delta R_c|<0.005$, which are below the present experimental sensitivity and therefore naturally `protect' the Standard Model prediction.

We are then left with four experimentally preferred charge assignments:
\begin{equation}
\begin{tabular}{c|rrr}
&$c_1$&$c_2$&$c_3$\\ \hline
2&0&$-{1\over2}$&1\\
5&$-{1\over2}$&0&1\\
11&$-{1\over2}$&1&$-{1\over2}$\\
12&$-{1\over2}$&$-{1\over2}$&1
\end{tabular}
\label{eq:prefs}
\end{equation}
We note that these charge assignments are unlike any that have been so far
considered in the literature. In fact, the popular assumption of a
generation-independent $\rm U'$ charge is violated explicitly in these four
successful cases. The constraints imposed by flipped SU(5) make that case
(13 in Table~\ref{Table0}) disfavored by experimental data. Note also
that the first two assignments imply
$\Delta\Gamma_{u\bar u}=\Delta\Gamma_{d\bar d}=0$ or
$\Delta\Gamma_{c\bar c}=\Delta\Gamma_{s\bar s}=0$, respectively. These perhaps
``unnatural" charge assignments occur quite naturally in the string model.
For future reference, we point out that in the string model $F_4$ is required
to contain the third-generation quarks in order to generate the top-quark
Yukawa coupling. From Eq.~(\ref{eq:Fcharges}) we see that we must have
$c_3=-{1\over2}$, which uniquely selects charge assignment (11)
in Eq.~(\ref{eq:prefs}). In this stringy preferred case we obtain the following
relation
\begin{equation}
\Delta R_b\approx 0.0042\,
\left({\Delta\Gamma_{\rm had}\over -3\,{\rm MeV}}\right)
\label{eq:predictions}
\end{equation}
(and also $\Delta R_c\approx -0.76\,\Delta R_b$).

Before proceeding, we should address the question of flavor-changing-neutral
currents (FCNC) that necessarily arise in our model because the generation
non-universal couplings of the $Z'$ to quarks spoil the GIM cancellation.
For a leptophobic $Z'$ the tree-level $Z'$-mediated contributions to $\Delta m_K,\Delta m_D$, and $\Delta m_B$ should provide the strongest constraints.
As usual, in obtaining the CKM matrix $V=U_LD^\dagger_L$ there is freedom of
how to choose $U_L$ and $D_L$ such that $V$ agrees with experimental observations. In Ref.~\cite{Frampton} it is shown that taking $U_L=1$, the
experimental value of $\Delta m_B$ forces our parameter $\delta$ to uninterestingly small values. Taking instead $D_L=1$ automatically satisfies
the $\Delta m_K$ and $\Delta m_B$ constraints, leaving only 
$\Delta m_D<10^{-13}\,{\rm GeV}$ to contend with. Following Ref.~\cite{Frampton} we find $\delta(\Delta m_D)\approx (7\times10^{-6}\,{\rm GeV})|\delta|^2 |(VgV^\dagger)_{12}|^2[f_D/(0.22\,{\rm GeV})]^2$, where we have used $V=U_L$. In our preferred charge assignment (11) we have
$g={\rm diag}(-{1\over2},1,-{1\over2})$, and we obtain $|(VgV^\dagger)_{12}|={1\over2}|V_{12}V^*_{22}|\approx{1\over10}$. Given the
inherent uncertainties in this kind of calculations, we conclude that our
above choice of $|\delta|\sim10^{-2}$ is consistent with FCNC constraints at the present time.

\section{Stringy flipped SU(5) leptophobia}
\label{sec:scenarios}
In the context of stringy flipped SU(5), we will consider four criteria in the
search for suitable U(1) gauge symmetries: they should be anomaly-free,
leptophobic, unbroken, and mixed. The anomaly-free requirement can only be
addressed in the context of a fundamental theory (such as string theory), where
anomaly-free symmetries are automatic (although they may need to be broken).
The (low-energy) leptophobia requirement is dictated by phenomenology and can
be enforced by symmetry, or achieved dynamically via mixing in the gauge
kinetic functions \cite{Holdom}. The new U(1) should remain unbroken after the
vacuum shifting at the Planck scale that is required to cancel the anomalous
$\rm U_A(1)$. Finally, since the new gauge boson must be leptophobic, it cannot
be  produced directly at LEP, and thus can affect LEP physics only via mixing
with the regular $Z$.

We seek the above properties in the context of a string-derived version of
the SU(5)$\times$U(1) model \cite{search} that has been shown to possess
various desirable properties regarding the vacuum energy, string unification,
the dynamical generation of all mass scales, the top-quark mass, and the strong
coupling \cite{goodies}. The complete gauge group of the model breaks up into
three identifiable pieces $G=G_{\rm obs}\times
G_{\rm hidden}\times G_{\rm U(1)}$, where $G_{\rm obs}=\rm SU(5)\times U(1)$,
$G_{\rm hidden}=\rm SU(4)\times SO(10)$, and $G_{\rm U(1)}=\rm U_1(1)\times
U_2(1)\times U_3(1)\times U_4(1)\times U_5(1)$. The 63 massless matter fields
in the model are listed in
Tables~\ref{Table1},\ref{Table2},\ref{Table3},\ref{Table4}, along with their
charges under $G_{\rm U(1)}$. Of note is the fact that $\rm
Tr\,U_{1,2,3,5}\not=0$, whereas $\rm Tr\,U_4=0$. These seemingly
anomalous symmetries are artifacts of the truncation of the full string
spectrum down to the massless sector. The low-energy effective theory is
correctly specified by first ``rotating" all the anomaly into a single
anomalous $\rm U_A\propto \sum_{i=1,2,3,5}\, [{\rm Tr}\,U_i]U_i$ \cite{HFM},
and then adding a one-loop correction to the D-term corresponding to $\rm U_A$:
$\rm D_A\to D_A+\epsilon M^2$, where $\epsilon=g^2{\rm Tr\,U_A}/192\pi^2$
\cite{DSW}.

The mass spectrum of all the states in the Tables can be obtained in a complicated procedure that takes into account the trilinear and non-renormalizable contributions to the superpotential and therefore to the masses and interactions \cite{search}. Such procedure does not have a unique outcome because the values of the vevs of the singlet fields in Table~\ref{Table2} are unknown (although constrained by the anomalous $\rm U_A$ cancellation conditions). The objective of this exercise is to obtain an electroweak-scale spectrum that resembles closely that of the MSSM. For instance, in Table~\ref{Table1} only one pair of Higgs pentaplets (their doublet components only) should survive to low energies. Below we will find
constraints on which states must belong to the low-energy sector of the theory.
We also note that the possible $\rm U'$ symmetries that we identify below
do not change the predictions of the string model for the superpotential Yukawa couplings, as these symmetries are derived along with the rest of the model. The problem of hierarchical fermion mass generation finds a possible resolution in string models. As is well known, several internal (world-sheet) symmetries severely restrict the allowed Yukawa couplings, typically predicting non-zero trilinear couplings for only the third family. Non-renormalizable terms usually contain the remaining Yukawa couplings, but these are naturally suppressed by powers of a small ratio that arises in the cancellation of the anomalous $\rm U_A$. This program has met with some success in the present model \cite{search}.

We find three scenarios containing possibly new light neutral gauge bosons.
\subsection{First scenario}
\label{sec:first}
The $\rm U_4(1)$ gauge symmetry is traceless (anomaly-free), the leptons
($\bar f_{2,3,5},\ell^c_{2,3,5}$) are not charged under it (leptophobic, see
Table~\ref{Table1}), and it
does not participate in the $\rm U_A(1)$ cancellation mechanism (unbroken).
Moreover, some of the quarks are charged under $\rm U_4(1)$. The problem is
that $\rm U_4(1)$ and $\rm U_Y(1)$ do not mix: the Higgs doublets that break
the electroweak symmetry are uncharged under $\rm U_4$ (see $h_i,\bar h_i$ in
Table~\ref{Table1}). The mixing via the gauge kinetic functions is not
operative, as one can easily verify that $\rm Tr\,(YU_4)=0$. This fact
``protects" the leptophobia, as otherwise the leptons would get their $\rm U_4$
charges shifted away from zero.

The only representations charged under $\rm U_4$ that may also contain quark
fields are $F_0,F_1$. Making this assignment also keeps $\rm U_4$ unbroken
through the SU(5)$\times$U(1) symmetry-breaking process at the GUT scale. The
$\rm U_4$ symmetry may be broken radiatively at low energies if the singlet
fields $\eta_{1,2},\bar\eta_{1,2}$, which solely carry the $\rm U_4$ charge
(see Table~\ref{Table2}), acquire suitable dynamical vacuum expectation values.

As such this $Z'$ will not affect LEP physics, as it does not obviously mix
with the $Z$. However, direct detection at hadron colliders via its hadronic
decays or parity-violating spin asymmetries is possible. If we assume that
$F_0,F_1$ contain first- and  second-generation fields ($F_4$ is expected to
contain the third generation fields \cite{search}) then the $Z'$ couplings to
quarks ($C'_{V,A}$) are as given in Eq.~(\ref{eq:charges}), but only for the
first- and second-generations. Consulting Table~\ref{Table1} one finds
$c_1=-c_2=\pm{1\over2}$ and $c_3=0$ which, in the event that some mechanism for
$Z$-$Z'$ mixing were to be found, would imply $\Delta\Gamma_{\rm had}=0$ and
$\Delta R_b=0$, leaving $R_b$ unshifted.

\subsection{Second scenario}
\label{sec:second}
There are three linear combinations of $\rm U_{1,2,3,5}$ that are orthogonal to
$\rm U_A=U_1-3U_2+U_3+2U_5$, and therefore traceless. A suitable basis is
provided by: $\rm U'_1=U_3+2U_5$, $\rm U'_2=U_1-3U_2$, $\rm
U'_3=3U_1+U_2+4U_3-2U_5$. Is there a linear combination of $\rm U'_{1,2,3}$
that is leptophobic? The leptons transform as $\bar
f_{2,5},\ell^c_{2,5}:(0,{3\over2},-{1\over2})$;
$\bar f_3,\ell^c_3:({3\over2},0,1)$, from which it follows that there is a
{\rm unique} leptophobic linear combination: $\rm U'\propto 2U'_1-U'_2-3U'_3
\propto U_1+U_3-U_5$. This symmetry is by construction anomaly-free and
leptophobic, and some of the Higgs pentaplets are charged under it ({\em i.e.},
mixed). The charges of all fields under $\rm U'$ are listed in the tables,
along with the two additional traceless combinations that can be chosen to
be $\rm U''=U_1-3U_2+U_3+2U_5$ and $\rm U'''=U_2+U_3+U_5$. From the tables we
see that only a very limited set of fields is uncharged under $\rm U'$,
\begin{equation}
F_2,\bar F_5,\Phi_{31},\bar\Phi_{31},T_1,T_2,T_3,D_3,D_7\ ,
\label{eq:vevs}
\end{equation}
and therefore their vevs will leave $\rm U'$ unbroken. The crucial question
is whether the usual D- and F-flatness conditions may be satisfied with such
a limited set of vevs. Moreover, this set of vevs generally breaks the hidden
sector gauge group. The feasibility of this scenario will be addressed in more
detail in Sec.~\ref{sec:revisited} below.

Let us assume that $\rm U'$ can indeed remain unbroken down to low energies.
One can verify that $\rm Tr\,(YU')=0$ (at least for an appropriately chosen
subset of light fields) and thus the $\rm U'$ charges are not shifted, and the
leptophobia is protected. Previous studies \cite{search} show that $F_4$ should
contain the third-generation quarks. Also, $F_2,\bar F_5$ should be engaged in
SU(5)$\times$U(1) breaking, as they are neutral under $\rm U'$, and therefore
$\bar F_4$ should contain the additional representations required for string
unification \cite{goodies}. It remains to determine the fate of $F_0$, $F_1$,
and $F_3$, two of which should be assigned to contain the first- and
second-generation quarks, whereas the third one will accompany $\bar F_4$.
Fortunately, the $R_b,R_c$ analysis of Sec.~\ref{sec:RbRc} provides an
important clue. Our assignment of $F_4$ entails $c_3=-{1\over2}$, and
Eq.~(\ref{eq:prefs}) then implies $c_1=-{1\over2}$ and $c_2=1$. This means that $F_3$ (which has $c=1$) should contain the second generation, whereas $F_0$ or $F_1$ (or a linear combination thereof) should contain the first generation.

\subsection{Third scenario}
\label{sec:third}
Since the unique leptophobic $\rm U'$ in the second scenario (in
Sec.~\ref{sec:second}) requires the hidden sector gauge group $G_{\rm hidden}$
to be broken in order to preserve D- and F-flatnes, one may consider giving up
a priori leptophobia, hoping to later be able to generate dynamic leptophobia
via $\rm U(1)$ mixing. This scenario may leave $G_{\rm hidden}$ and the desired
$\rm U(1)$ unbroken, and still preserve flatness. To this end we determine
all possible linear combinations of $\rm U'_1,U'_2,U'_3$ (or $\rm U',U'',U'''$)
that are uncharged under more than one singlet field (in Table~\ref{Table2}),
with the idea that vevs from this subset of fields should suffice to preserve
flatness. We find four such linear combinations:
\begin{center}
\begin{tabular}{lcl}
$\rm U_a=U_1+U_2+2U_3$&\qquad\qquad&
${\cal S}_a=\{\Phi_{12},\bar\Phi_{12},\phi^+,\bar\phi^+,\phi^-,\bar\phi^-,
\phi_{3,4},\bar\phi_{3,4}\}$\\
$\rm U_b=U_1+U_2+U_3-2U_5$&&
${\cal S}_b=\{\Phi_{12},\bar\Phi_{12},\Phi_{23},\bar\Phi_{23},
\phi_{3,4},\bar\phi_{3,4}\}$\\
$\rm U_c=U_1+U_2-U_3-6U_5$&&
${\cal S}_c=\{\Phi_{12},\bar\Phi_{12},\phi_{45},\bar\phi_{45},
\phi_{3,4},\bar\phi_{3,4}\}$\\
$\rm U_d=U_1-3U_2+U_3+2U_5$&&
${\cal S}_d=\{\Phi_{31},\bar\Phi_{31},\phi^-,\bar\phi^-,
\phi_{45},\bar\phi_{45}\}$
\end{tabular}
\end{center}
listed along with the singlet fields that are uncharged under them. It is easy
to show that the D-flatness conditions specific to this model (as given
in Ref.~\cite{search}) cannot be satisfied by vevs belonging to the
subsets ${\cal S}_a$, ${\cal S}_b$, or ${\cal S}_c$. Subset ${\cal S}_d$ may
work. As one can verify, the leptons ($\bar f_{2,3,5},\ell^c_{2,3,5}$) couple
universally to $\rm U_d$, via a charge that would need to be shifted down to
zero by the dynamical mixing mechanism (which is available as $\rm Tr\,
(YU_d)\not=0$). Unfortunately, universal leptophobia is not achievable because
the dynamical shifts would be proportional to the hypercharge of the leptons,
and thus would not be universal.

\section{Second scenario revisited}
\label{sec:revisited}
There are several questions that need to be addressed in order to insure that
a U(1) symmetry at the string scale indeed becomes the $\rm U'$ symmetry
with the desired properties at the electroweak scale. One must insure that
this symmetry remains unbroken in the anomalous $\rm U_A$ cancellation
mechanism, and upon SU(5)$\times$U(1) symmetry breaking. After these hurdles
have been cleared, on the way down to low energies the $g_{Z'}$ gauge
coupling must evolve properly, and the gauge symmetry must be dynamically
broken near the electroweak scale. These two effects give rise to the
phenomenological parameter $\delta$ in Eq.~(\ref{eq:delta}), which must obtain
a value in the range of interest ($|\delta|\sim10^{-2}$). Addressing all these
questions requires a very detailed study of the string model, which does not
appear warranted at this point, given the uncertainty in the experimental
measurements. Nonetheless, we will attempt to outline the main issues to be
confronted in such a future study.

First of all, in the anomalous $\rm U_A$ cancellation mechanism three gauge
symmetries must be broken $\rm U_A,U'',U'''$, but $\rm U'$ must remain
unbroken. Moreover, it does not help to have some linear combination involving
$\rm U'$ to remain unbroken, as $\rm U'$ will be mixed and leptophobia would
likely be lost. The vector-boson mass matrix is given by
${\cal M}^2_{\alpha\beta}\propto \sum_a (\partial D_\alpha/\partial\phi_a)
(\partial D_\beta/\partial\phi_a)^*$, where
$D_\alpha=g\sum_i q^\alpha_i\phi_i\phi^*_i$, and $\alpha,\beta=\rm
U_A,U',U'',U'''$. The resulting $4\times4$ matrix must have zeroes for all
entries involving $\rm U'$:
\begin{equation}
{\cal M}^2_{\rm U'U'}\propto\sum_a|\partial D_{\rm U'}/\partial\phi_a|^2=
g^2\sum_a (q'_a)^2|\phi_a|^2=0\ ,
\label{eq:condition}
\end{equation}
is a necessary and sufficient condition, which implies that only fields
uncharged under $\rm U'$ (those in Eq.~(\ref{eq:vevs})) should be allowed to
acquire vevs. In this case the D-flatness conditions become:
\begin{eqnarray}
D_A:&&{\textstyle{1\over2}}(|V_2|^2-|\bar V_5|^2)+
{\textstyle{5\over2}}(|T_1|^2+|T_2|^2+|T_3|^2+|D_3|^2-|D_7|^2)-x_{31}+\epsilon M^2=0
\nonumber\\
&&\label{eq:DA}\\
D':&&0=0
\label{eq:D'}\\
D'':&&{\textstyle{3\over2}}(|V_2|^2-|\bar V_5|^2)=0
\label{eq:D''}\\
D''':&&-{\textstyle{1\over2}}(|V_2|^2-|\bar V_5|^2)
+{\textstyle{1\over2}}(|T_1|^2+|T_3|^2
+|D_3|^2-|D_7|^2)-|T_2|^2-x_{31}=0
\label{eq:D'''}
\end{eqnarray}
where we have defined $x_{31}=|\Phi_{31}|^2-|\bar\Phi_{31}|^2$, and
$\epsilon=g^2{\rm Tr\,U_A}/192\pi^2>0$. These equations have the solution
\begin{eqnarray}
|V_2|&=&|\bar V_5|\\
x_{31}&=&-{\textstyle{1\over4}}\epsilon M^2 -{\textstyle{15\over8}}|T_2|^2<0\\
|T_1|^2+|T_3|^2 +|D_3|^2-|D_7|^2&=&-{\textstyle{1\over2}}\epsilon M^2
-{\textstyle{7\over4}}|T_2|^2<0
\end{eqnarray}
Note that the last two equations imply $\langle\bar\Phi_{31}\rangle\not=0$ and $\left\langle D_7\right\rangle\not=0$.

One also has to consider the SU(5)$\times$U(1), SU(4), and SO(10) flatness
conditions. The first one implies $|V_2|=|\bar V_5|$, which is satisfied
automatically when the $\rm D''$ condition is satisfied (Eq.~(\ref{eq:D''})).
The SU(4) and SO(10) conditions are given by \cite{ART}
\begin{equation}
D_{\rm SU(4)}:\quad \sum_{a=3,7} D^*_a\,\tau^\alpha\, D_a=0\,;\qquad
D_{\rm SO(10)}:\quad \sum_{a=1,2,3} T^*_a\,\lambda^A\, T_a=0\ ,
\label{eq:SU4SO10}
\end{equation}
where $\tau^\alpha,\lambda^A$ are the generators of SU(4) and SO(10)
respectively. In the basis in which the generators are given by the imaginary
antisymmetric matrices, the $D_a\,(T_a)$ are represented by 6-vectors
(10-vectors). Specific choices for the vevs of these representations will
determine the pattern of SU(4) and SO(10) symmetry breaking. It is worth
noticing and easy to verify that these flatness conditions are automatically
satisfied for $D_a,T_a$ vevs that are real \cite{ART}. However such restriction
is not necessary; relaxing it allows situations where, {\em e.g.}, $|D|^2=D\cdot D^*\not=0$, while $D\cdot D=0$.

The standard F-flatness conditions require that $\partial W/\partial\phi_a=0$
at the minimum of the scalar potential. Considering only the cubic
contributions to $W$ \cite{search}, we obtain
\begin{equation}
\begin{tabular}{lll}
${\partial W\over\partial \Phi_{31}}=T_1\cdot T_1+T_3\cdot T_3+D_3\cdot
D_3\qquad$
&${\partial W\over\partial T_1}=T_1\Phi_{31}\qquad$
&${\partial W\over\partial D_3}=D_3\Phi_{31}$\\
&&\\
${\partial W\over\partial \bar\Phi_{31}}=D_7\cdot D_7$
&${\partial W\over\partial T_3}=T_3\Phi_{31}$
&${\partial W\over\partial D_7}=D_7\bar\Phi_{31}$
\end{tabular}
\label{eq:Fflat}
\end{equation}
As such, the $\partial W/\partial D_7$ condition appears to be in conflict with
the D-flatness requirement of $\bar\Phi_{31},D_7\not=0$. Inclusion of
non-renormalizable contributions to the superpotential may resolve this
impasse.

In the standard string unification picture, one would expect the new gauge
coupling $g_{Z'}$ to evolve from low energies up to the string scale, where
it will meet the other gauge couplings. In our case the proper normalization
of $\rm U'$ is well defined from the requirement that all massless fields
have conformal dimension 1: $\rm U'\to U'/\sqrt{3}$. The gauge coupling evolves
according to the beta function $b'={1\over3}\rm Tr\, (Q')^2$, where $Q'$ are
the charges that appear in Tables~\ref{Table1}--\ref{Table4}. Defining the
MSSM matter content as consisting of the quark and lepton fields in
$F_{0,3,4},\bar f_{2,3,5},\ell^c_{2,3,5}$, and the Higgs doublets in $h_1,\bar
h_{45}$ \cite{search}, we find $b'={16\over3}$, which is smaller than the
traditional $b_Y={33\over5}$. At the scale where the intermediate $F_1,\bar
F_4$ states become excited, we find $b'={41\over6}$, which is very close to
$b_Y$. Therefore, at least up to the unification scale the evolution of $\rm
U'$ is reasonable. The evolution up to the string scale requires a detailed
understanding of the spectrum at such mass scales.

One also needs to worry about the $\rm U'$ symmetry breaking mechanism. Simply
assuming that the usual Higgs doublets, which are charged under $\rm U'$, will
effect the symmetry breaking will lead to a much-too-large mixing angle
$\theta$ \cite{CL}. One must resort to a singlet field acquiring a vev
radiatively \cite{EENZ}, much in the same spirit as in the original flipped
SU(5) picture \cite{faspects}. This possibility looks promising, as in this
scenario there are many singlet fields that do not acquire vevs at the string
scale. In this case one expects $\delta\sim\theta\sim M^2_Z/M^2_{Z'}$, and the
phenomenological requirement (from fits to electroweak data, including $R_b$)
of $\delta\sim10^{-2}$ naively implies $M_{Z'}\sim 10M_Z$, although ${\cal
O}(1)$ factors may alter this relation either way.

\section{Experimental prospects}
\label{sec:exp}
In this section we address some of the experimental consequences of the $Z'$
gauge bosons proposed above. We concentrate on the second scenario in
Sec.~\ref{sec:scenarios} (case (11) in Eq.~(\ref{eq:prefs})), as this is the one that may shift $R_b$ in an interesting way. (The predicted correlation between $\Delta \Gamma_{\rm had}$ and $\Delta R_b$ has been given in Eq.~(\ref{eq:predictions}) above.)
This analysis will be sketchy. More precise calculations will be warranted once some as-yet-unknown parameters ({\em i.e.}, $\theta,g_{Z'},M_{Z'}$) become available. In this spirit we neglect the effects of $Z$-$Z'$ mixing and concentrate on the direct production and decay of our leptophobic $Z'$ at hadron colliders. The only variables in this case will be the strength of the $\rm U'$ gauge coupling ($g_{Z'}$) and the mass of the $Z'$. 

\subsection{$Z'$ width and branching ratios}
\label{sec:width}
The $Z'$ partial width into quarks can be expressed as \cite{Barger}
\begin{equation}
\Gamma(Z'\to q\bar q)={\Gamma_0\over M_Z}
\left({g_{Z'}\over g_Z}\right)^2({\textstyle{1\over3}})
\left[C^{'2}_V+C^{'2}_A\right]M_{Z'}\ ,
\label{eq:width}
\end{equation}
where $\Gamma_0/M_Z=G_F M^2_Z/(2\sqrt{2}\pi)\approx0.011$, and the $C'_V,C'_A$
couplings can be obtained from Eq.~(\ref{eq:charges}) for the preferred
charge assignment in Eq.~(\ref{eq:prefs}) ({\em i.e.}, case (11)). The explicit factor of $1\over3$ in Eq.~(\ref{eq:width}) has been inserted to
normalize the $\rm U'$ charges properly, as discussed in Sec.~\ref{sec:revisited}. The relevant couplings for the case of interest
are:
\begin{equation}
\begin{tabular}{c|rrc}
&$C'_V$&$C'_A$&$B(Z'\to q\bar q)$\\ \hline
$u$&$-{1\over2}$&$-{1\over2}$&$1\over18$\\
$d$&$-1$&0&$1\over9$\\
$c$&1&1&$2\over9$\\
$s$&2&0&$4\over9$\\
$t$&$-{1\over2}$&$-{1\over2}$&$1\over18$\\
$b$&$-1$&0&$1\over9$
\end{tabular}
\label{eq:CVA}
\end{equation}
The branching ratios have been calculated by neglecting the top-quark mass
({\em i.e.}, for $M_{Z'}\gg 2m_t$); they are only slightly increased if the top-quark does not fully contribute. Note the strong preference for light-quark decays. The total $Z'$ width, neglecting the top-quark mass, is given by \begin{equation}
{\Gamma_{Z'}\over M_{Z'}}\approx 0.033 \left({g_{Z'}\over g_Z}\right)^2\ ,
\label{eq:TotalWidth}
\end{equation}
which shows that our leptophobic $Z'$ is expected to be narrow (assuming that
$g_{Z'}\sim g_Z$), and therefore amenable to the standard searches for $Z'$ to
dijets at hadron colliders \cite{UA2,Harris}. (If supersymmetric particle
decays are allowed, the width will increase although still remain relatively
narrow.)

\subsection{$Z'$ production and experimental limits}
\label{sec:production}
Direct channel production of a narrow $Z'$ at hadron colliders has a
parton-level cross section given by \cite{Barger}
\begin{equation}
\widehat\sigma(q\bar q\to Z')=K\, {\textstyle{4\pi^2\over3}}\, {\Gamma_0\over M_Z}\left({g_{Z'}\over g_Z}\right)^2({\textstyle{1\over3}})\left[C^{'2}_V+C^{'2}_A\right]
\delta(\widehat s-M^2_{Z'})\ ,
\label{eq:sigma}
\end{equation}
with an estimated K-factor of $K\approx1.3$ \cite{Barger}. For our present
purposes it should suffice to determine the ratio of our $Z'$ cross section
to that obtained assuming a $Z'$ with Standard Model couplings to quarks (as
traditionally assumed in the experimental literature). These ratios are
\begin{equation}
{\widehat\sigma(u\bar u\to Z')\over \widehat\sigma(u\bar u\to Z')_{\rm SM}}
\approx 0.58 \left({g_{Z'}\over g_Z}\right)^2\ ;\qquad
{\widehat\sigma(d\bar d\to Z')\over \widehat\sigma(d\bar d\to Z')_{\rm SM}}
\approx 0.90\left({g_{Z'}\over g_Z}\right)^2\ .
\label{eq:ratios}
\end{equation}
To estimate the impact of
present experimental searches for $Z'$ into dijets, one could average the
contributions from up and down quarks (which gives 0.75 average coefficient) and then multiply by $B(Z'\to jj)/B(Z'\to jj)_{\rm SM}\approx1.4$,
\begin{equation}
\sigma(p\bar p\to Z'\to jj)\approx 
\left({g_{Z'}\over g_Z}\right)^2 \sigma(p\bar p\to Z'\to jj)_{\rm SM} \ .
\label{eq:approx}
\end{equation}
Assuming $g_{Z'}=g_Z$, one can study the exclusion plots obtained by UA2
\cite{UA2} and CDF \cite{Harris} in their searches for $Z'\to jj$. We conclude
that the UA2 lower bound of $M_{Z'}>260\,{\rm GeV}$ is applicable, whereas
CDF does not impose any further constraints. We note that even the UA2 lower
bound may be easily evaded should $g_{Z'}<g_Z$.

\subsection{$Z'$ effects on top-quark production}
\label{sec:ZprimeTop}
\begin{figure}[p]
\vspace{6in}
\includegraphics{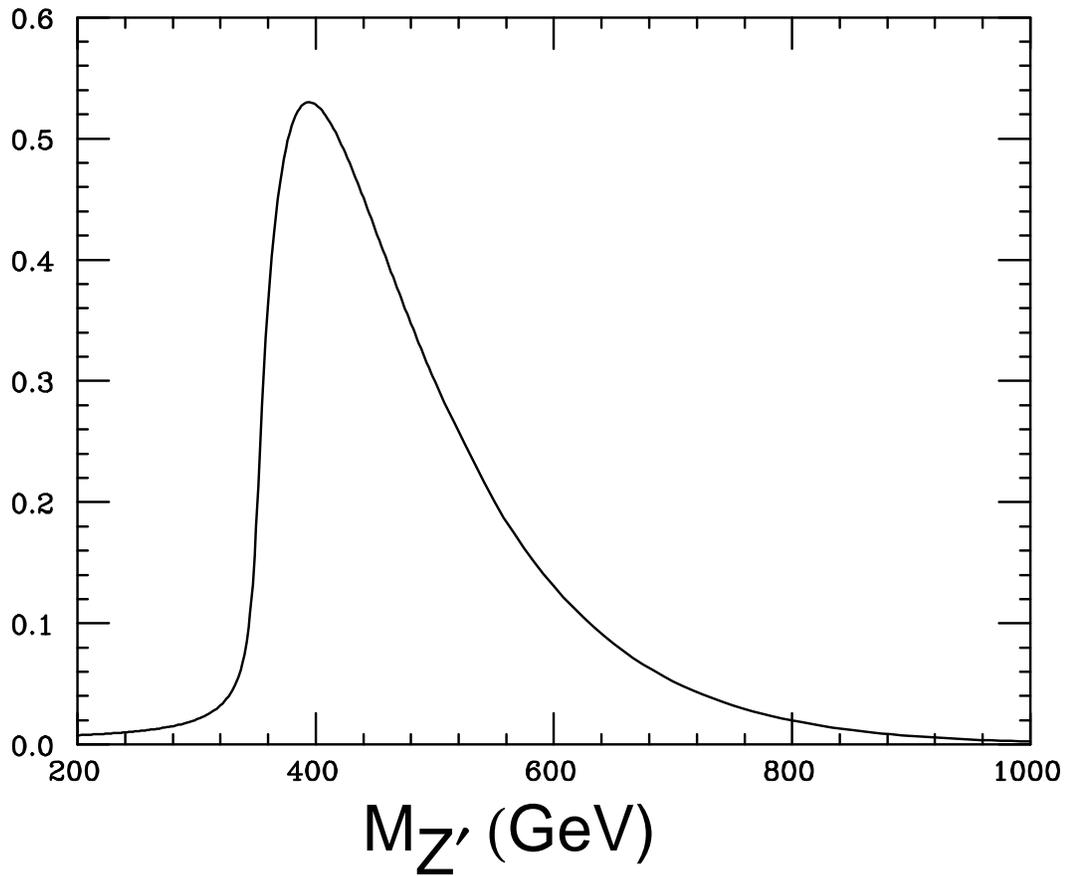}
\caption{The calculated cross section $\sigma(p\bar p\to Z'\to t\bar t)$
versus $M_{Z'}$ (for $g_{Z'}/g_Z=1$ and $m_t=175\,{\rm GeV}$) at the Tevatron
for the charge assignment of interest. (The cross section scales with
$(g_{Z'}/g_Z)^4$.) For reference 
$\sigma^{\rm SM}_{t\bar t}=4.75\pm0.6\,{\rm pb}$.}
\label{fig:Zprime}
\end{figure}

Our leptophobic $Z'$ will contribute to the top-quark cross section at the
Tevatron. The tree-level $q\bar q$ parton-level cross sections are given by
\cite{Stirling}
\begin{eqnarray}
\widehat\sigma(q\bar q\to g\to t\bar t)&=&{4\pi\alpha_s^2\over27\widehat s}\,
\beta\,(3-\beta^2)\ ,
\label{eq:old}\\
\widehat\sigma(q\bar q\to Z'\to t\bar t)&=&{\textstyle{4\pi\over3}}
\left({\Gamma_0\over M_Z}\right)^2
\left({g_{Z'}\over g_Z}\right)^4
{\widehat s\over(\widehat s-M^2_{Z'})^2+(\widehat s\Gamma_{Z'}/M_{Z'})^2}
\nonumber\\
&\times&({\textstyle{1\over3}})^2 \left[C^{'2}_{V_q}+C^{'2}_{A_q}\right]
 \left[{\textstyle{\beta\over2}}(3-\beta^2)C^{'2}_{V_t}
+\beta^3C^{'2}_{A_t}\right]\ ,
\label{eq:new}
\end{eqnarray}
where $\beta^2=1-4m^2_t/\widehat s$. Since $\alpha_s/(\Gamma_0/M_Z)\approx10$,
the $Z'$ contribution is usually negligible compared with the QCD
contribution. There are two exceptions: (i) either the $C_{V,A}$ charges are
large, or (ii) resonant production ({\em i.e.}, $\widehat s\approx M^2_{Z'}$) is possible. The model of Ref.~\cite{Altarelli} satisfies the first exception
but not the second one, as they fix $M_{Z'}=1\,{\rm TeV}$, while at the Tevatron top-quark production occurs near threshold ({\em i.e.}, $\sqrt{\widehat s}\sim 2m_t$).
Our model does not satisfy the first exception, but it may yield observable
effects by satisfying the second exception if $M_{Z'}\sim 400\,{\rm GeV}$. This
expectation has been verified explicitly by integrating $\widehat\sigma$ over
the parton distribution functions (taken from Ref.~\cite{MT}). The result
for $\sqrt{s}=1.8\,{\rm TeV}$, $m_t=175\,{\rm GeV}$, and $g_{Z'}/g_Z=1$ is shown in Fig.~\ref{fig:Zprime} as a function of $M_{Z'}$ for the charge assignment of interest; a K factor of $K=1.3$ has been applied
\cite{Stirling}. Current experimental precision on $\sigma_{t\bar t}$ does not
yet allow one to conclusively exclude any such ($\sim 0.5\,{\rm pb}$) shifts. For reference, the corresponding Standard Model cross section is $\sigma^{\rm SM}_{t\bar t}=4.75\pm0.6\,{\rm pb}$ \cite{Berger}, showing that the $Z'$ effects may exceed the present theoretical uncertainty in the Standard Model prediction and therefore be in principle discernible in future runs at the Tevatron. Of course, the possible $Z'$ effects on the top-quark cross section are constrained by the allowed values of $M_{Z'}$ and $g_{Z'}/g_Z$ that result from the usual $Z'\to jj$ searches discussed above.

\subsection{Parity-violating spin asymmetries at RHIC}
\label{sec:RHIC}
A novel way in which our leptophobic $Z'$ may be detected takes advantage of
its maximal parity-violating couplings to up-type quarks (and parity-conserving
couplings to down-type quarks). This property ($C^u_L=c,C^u_R=0$;
$C^d_L=C^d_R=c$) depends solely on the underlying flipped SU(5) symmetry, and
is not affected by the details of the string model. The RHIC Spin Collaboration
\cite{RSC} intends to measure various spin asymmetries when RHIC starts
colliding polarized protons at center-of-mass energies as large as
$\sqrt{s}=500\,{\rm GeV}$, with an integrated luminosity in excess of 
$1\,{\rm fb}^{-1}$. The spin asymmetry of greatest sensitivity in our case is defined by \cite{BGS}
\begin{equation}
A^{\rm PV}_{\rm LL}={d\sigma_{--}-d\sigma_{++}\over d\sigma_{--}+d\sigma_{++}}
\ ,
\label{eq:APVLL}
\end{equation}
where $d\sigma_{\lambda_1\lambda_2}$ represents the cross section for
scattering of protons of given helicities producing a jet
($p^{\lambda_1}p^{\lambda_2}\to j+X$). It has been shown that at RHIC energies
the dominant contribution to the asymmetry comes from the interference between
gluon exchange and $Z'$ exchange in the $t$-channel scattering of quarks of the
same flavor. The parton-level asymmetry for a given quark flavor is then
proportional to \cite{TV}
\begin{equation}
T^{--}_{gZ'}-T^{++}_{gZ'}=(C^{'2}_L-C^{'2}_R)
{\textstyle{8\over9}}\alpha_s \alpha_{Z'}\,
\widehat s^2\, {\rm Re}\,
\left({1\over \widehat t D_{Z'}(\widehat t)}+
{1\over \widehat u D_{Z'}(\widehat u)}\right)\ ,
\label{eq:amplitude}
\end{equation}
where $\alpha_{Z'}=g^2_{Z'}/4\pi$ and
$D_{Z'}(x)=x-M^2_{Z'}+iM_{Z'}\Gamma_{Z'}$. It is then clear that down-type
quarks will not contribute to the asymmetry (as they have $C'_L=C'_R$), whereas
up-type quarks will contribute maximally, yielding (as $\widehat t,\widehat
u<0$)
\begin{equation}
A^{\rm PV}_{\rm LL}>0\ ,
\label{eq:sign}
\end{equation}
once the integration over polarized quark distribution functions is performed.
The Standard Model QCD-electroweak contribution to $A^{\rm PV}_{\rm LL}$ is
also positive. Examples of observable parity-violating spin asymmetries have
been displayed in Ref.~\cite{TV} for the case of $M_{Z'}=1\,{\rm TeV}$, and require
rather large couplings (as proposed in Ref.~\cite{Altarelli}). Such large
couplings are not available in our present model, but an observable asymmetry
may still be present for lighter $Z'$ masses. Examining
Eq.~(\ref{eq:amplitude}), it appears that for $M_{Z'}\gg \sqrt{\widehat s}\sim
100\,{\rm GeV}$ one obtains an $\sim 1/M^2_{Z'}$ dependence on the amplitude (and
therefore on the asymmetry).

The above qualitative analysis should motivate detailed studies of $A^{\rm
PV}_{\rm LL}$ at RHIC. Of particular  interest should be the case of a
leptophobic $Z'$ in the context of flipped SU(5), where the only unknowns are
$M_{Z'}$ and the product $(C^u_L\,g_{Z'})^2$ (which appears as an overall
constant), and the sign of the asymmetry is predicted to be positive. Note also
that the $Z$-$Z'$ mixing considered above plays no role in the prediction for $A^{\rm PV}_{\rm LL}$, and therefore even if such mixing is eventually found not to be relevant, parity-violating spin asymmetries at RHIC would still be of great interest in probing models of physics at very high energies.

\section{Conclusions}
\label{sec:conclusions}
We have addressed the possible existence of leptophobic $Z'$ gauge bosons
in consistent unified theories, in particular in the context of an underlying
flipped SU(5) gauge group. Leptophobia in this case is natural, as quarks are
largely split from leptons in the SU(5) representations. In contrast,
traditional unified gauge groups seem to require a dynamical mechanism for
generating leptophobia at the electroweak scale. Our leptophobic $Z'$ possesses
distinct couplings to quarks, violating parity maximally in the up-quark
sector, and not at all in the down-quark sector. Moreover, string-based $Z'$
charge assignments lead to scenarios where $R_b$ is shifted in the direction indicated experimentally, while keeping $\Gamma_{\rm had}$ and $R_c$ essentially unchanged. We have considered the origins of such phenomenologically desirable $Z'$ in the context of string-derived flipped SU(5), and identified three possible scenarios. One of them is particularly compelling, and has been studied further.

We have also determined some basic properties of our leptophobic $Z'$ gauge
boson, such as its total width, branching ratios, and production cross section.
Current experimental limits from $Z'$ to dijet searches may be applicable
if the $\rm U'$ coupling is comparable to the weak coupling. We have also
studied the effect of $Z'$ exchange on the top-quark cross section. Also, the
parity-violating couplings to up-type quarks have the potential of yielding
observable spin asymmetries in polarized $pp$ scattering at RHIC.

It is important to realize that even if $Z$-$Z'$ mixing is found not to be
relevant, leptophobic $Z'$ gauge bosons may still be predicted by string models (unmixed or negligible mixed with the $Z$), and their existence should be probed experimentally in all possible ways. In the context of flipped SU(5), spin asymmetries at RHIC may be particularly sensitive probes.

The next step along these lines should include a more detailed study of the
preferred string scenario, including non-renormalizable terms in the
superpotential and soft-supersymmetry-breaking terms. When a more clear picture
of the effective theory below the string scale emerges, one should attempt
a full dynamical evolution of the model down to low energies, paying particular
attention to the radiative breaking of the $\rm U'$ symmetry via vevs of
singlet fields. This crucial step will determine $M_{Z'}$, $g_{Z'}$, and
$\theta$, as well as the spectrum of the new light degrees of freedom that
accompany the singlet field.

\section*{Acknowledgments}
J. L. would like to thank Geary Eppley, Teruki Kamon, and Jay Roberts for very
useful discussions.
The work of J.~L. has been supported in part by DOE grant DE-FG05-93-ER-40717.
The work of D.V.N. has been supported in part by DOE grant DE-FG05-91-ER-40633.

\newpage
\begin{table}[p]
\caption{The 13 possible assignments of $\rm U'$ charges to the three
generations, along with the fractional changes in $\Gamma_{\rm had}$, $R_b$,
and $R_c$ (in units of $\delta$). For $\delta=0.01$ we also display the shift
in $\Gamma_{\rm had}$ in MeV, and the actual shifts in $R_b$ and $R_c$.}
\label{Table0}
\begin{center}
\begin{tabular}{|r|rrr|rrr|rrr|}\hline
&$c_1$&$c_2$&$c_3$&$\Delta\Gamma_{\rm had}/\Gamma_{\rm had}$&$\Delta R_b/R_b$
&$\Delta R_c/R_c$&$\Delta\Gamma_{\rm had}$&$\Delta R_b$&$\Delta R_c$\\ \hline
1& $   0$&$  -{1\over2}$&$  -{1\over2}$&$   1.00$&$    .87$&$    .06$&$
17.5$&$    .0019$&$    .0001$\\
2& $   0$&$  -{1\over2}$&$  1$&$   -.23$&$  -3.52$&$   1.30$&$     -4.0$&$
-.0076$&$    .0022$\\
3& $   0$&$  1$&$  -{1\over2}$&$   -.78$&$   2.65$&$  -1.36$&$    -13.5$&$
.0057$&$   -.0023$\\
4& $  -{1\over2}$&$   0$&$  -{1\over2}$&$   1.00$&$    .87$&$  -1.00$&$
17.5$&$    .0019$&$   -.0017$\\
5& $  -{1\over2}$&$   0$&$  1$&$   -.23$&$  -3.52$&$    .23$&$     -4.0$&$
-.0076$&$    .0004$\\
6& $  1$&$   0$&$  -{1\over2}$&$   -.78$&$   2.65$&$    .78$&$    -13.5$&$
.0057$&$    .0013$\\
7& $  -{1\over2}$&$  -{1\over2}$&$   0$&$   1.19$&$  -1.19$&$   -.12$&$
20.7$&$   -.0026$&$   -.0002$\\
8& $  -{1\over2}$&$  1$&$   0$&$   -.59$&$    .59$&$  -1.54$&$    -10.3$&$
.0013$&$   -.0027$\\
9& $  1$&$  -{1\over2}$&$   0$&$   -.59$&$    .59$&$   1.66$&$    -10.3$&$
.0013$&$    .0029$\\
10& $  1$&$  -{1\over2}$&$  -{1\over2}$&$   -.18$&$   2.05$&$   1.25$&$
-3.2$&$    .0044$&$    .0021$\\
11& $  -{1\over2}$&$  1$&$  -{1\over2}$&$   -.18$&$   2.05$&$  -1.95$&$
-3.2$&$    .0044$&$   -.0034$\\
12& $  -{1\over2}$&$  -{1\over2}$&$  1$&$    .36$&$  -4.11$&$    .70$&$
6.3$&$   -.0089$&$    .0012$\\
13&  $  -{1\over2}$&$  -{1\over2}$&$  -{1\over2}$&$   1.60$&$    .27$&$
-.53$&$     27.9$&$    .0006$&$   -.0009$\\ \hline
\end{tabular}
\end{center}
\end{table}
\clearpage

\begin{table}[p]
\caption{The observable sector massless matter fields and their transformation
properties under $G_{\rm U(1)}$. Under SU(5)$\times$U(1) these fields transform
as $F=(10,{1\over2})$, $\bar f=(\bar 5,-{3\over2})$, $\ell^c=(1,{5\over2})$,
$h=(5,-1)$, and $\bar h=(\bar 5,1)$. Also indicated are the charges under
$\rm U_A$ and three orthogonal linear combinations of interest ($\rm
U',U'',U'''$).}
\label{Table1}
\smallskip
\begin{center}
\small
\begin{tabular}{|l|rrrrr|rrrr|}\hline
&$\rm U_1$&$\rm U_2$&$\rm U_3$&$\rm U_4$&$\rm U_5$&$\rm U_A$
&$\rm U'$&$\rm U''$&$\rm U'''$\\
\hline
$F_0$&$-{1\over2}$&0&0&$-{1\over2}$&0&${3\over2}$&$-{1\over2}$&$-{1\over2}$&0\\
$F_1$&$-{1\over2}$&0&0&${1\over2}$&0&${3\over2}$&$-{1\over2}$&$-{1\over2}$&0\\
$F_2$&0&$-{1\over2}$&0&0&0&${1\over2}$&0&${3\over2}$&$-{1\over2}$\\
$F_3$&0&0&${1\over2}$&0&$-{1\over2}$&${3\over2}$&1&$-{1\over2}$&0\\
$F_4$&$-{1\over2}$&0&0&0&0&${3\over2}$&$-{1\over2}$&$-{1\over2}$&0\\
$\bar F_4$&${1\over2}$&0&0&0&0&$-{3\over2}$&${1\over2}$&${1\over2}$&0\\
$\bar F_5$&0&${1\over2}$&0&0&0&$-{1\over2}$&0&$-{3\over2}$&${1\over2}$\\
$\bar f_2,\ell^c_2$&0&$-{1\over2}$&0&0&0&${1\over2}$&0
&${3\over2}$&$-{1\over2}$\\
$\bar f_3,\ell^c_3$&0&0&${1\over2}$&0&${1\over2}$&${1\over2}$&0&${3\over2}$&1\\
$\bar f_5,\ell^c_5$&0&$-{1\over2}$&0&0&0&${1\over2}$&0
&${3\over2}$&$-{1\over2}$\\
$h_1$&1&0&0&0&0&$-3$&1&1&0\\
$\bar h_1$&$-1$&0&0&0&0&3&$-1$&$-1$&0\\
$h_2$&0&1&0&0&0&$-1$&0&$-3$&1\\
$\bar h_2$&0&$-1$&0&0&0&1&0&3&$-1$\\
$h_3$&0&0&1&0&0&2&1&1&1\\
$\bar h_3$&0&0&$-1$&0&0&$-2$&$-1$&$-1$&$-1$\\
$h_{45}$&$-{1\over2}$&$-{1\over2}$&0&0&0&2&$-{1\over2}$&1&$-{1\over2}$\\
$\bar h_{45}$&${1\over2}$&${1\over2}$&0&0&0&$-2$&${1\over2}$&$-1$&${1\over2}$\\
\hline
\end{tabular}
\caption{The singlet fields and their transformation properties
under $G_{\rm U(1)}$, $\rm U_A$, and three orthogonal linear combinations of
interest ($\rm U',U'',U'''$).}
\label{Table2}
\smallskip
\begin{tabular}{|l|rrrrr|rrrr|}\hline
&$\rm U_1$&$\rm U_2$&$\rm U_3$&$\rm U_4$&$\rm U_5$&$\rm U_A$
&$\rm U'$&$\rm U''$&$\rm U'''$\\
\hline
$\Phi_{12}$&$-1$&1&0&0&0&2&$-1$&$-4$&1\\
$\bar\Phi_{12}$&1&$-1$&0&0&0&$-2$&1&4&$-1$\\
$\Phi_{23}$&0&$-1$&1&0&0&3&1&4&0\\
$\bar\Phi_{23}$&0&1&$-1$&0&0&$-3$&$-1$&$-4$&0\\
$\Phi_{31}$&1&0&$-1$&0&0&$-5$&0&0&$-1$\\
$\bar\Phi_{31}$&$-1$&0&1&0&0&5&0&0&1\\
$\phi_{45}$&${1\over2}$&${1\over2}$&1&0&0&0&${3\over2}$&2&${3\over2}$\\
$\bar\phi_{45}$&$-{1\over2}$&$-{1\over2}$&$-1$&0&0&0&$-{3\over2}$
&$-2$&$-{3\over2}$\\
$\phi^+$&${1\over2}$&$-{1\over2}$&0&0&1&$-2$&$-{1\over2}$&4&${1\over2}$\\
$\bar\phi^+$&$-{1\over2}$&${1\over2}$&0&0&$-1$&2&${1\over2}$&$-4$
&$-{1\over2}$\\$\phi^-$&${1\over2}$&$-{1\over2}$&0&0&$-1$&0&${3\over2}$&0
&$-{3\over2}$\\
$\bar\phi^-$&$-{1\over2}$&${1\over2}$&0&0&1&0&$-{3\over2}$&0&${3\over2}$\\
$\phi_{3,4}$&${1\over2}$&$-{1\over2}$&0&0&0&$-1$&${1\over2}$&2&$-{1\over2}$\\
$\bar\phi_{3,4}$&$-{1\over2}$&${1\over2}$&0&0&0&1&$-{1\over2}$&$-2$&$1\over2$\\
$\eta_{1,2}$&0&0&0&1&0&0&0&0&0\\
$\bar\eta_{1,2}$&0&0&0&$-1$&0&0&0&0&0\\
$\Phi_{0,1,3,5}$&0&0&0&0&0&0&0&0&0\\
\hline
\end{tabular}
\end{center}
\end{table}
\clearpage

\begin{table}[p]
\caption{The hidden SO(10) decaplets ({\bf10}) $T_i$ fields and their
transformation properties under $G_{\rm U(1)}$. Also indicated are the charges
under $\rm U_A$ and three orthogonal linear combinations of interest ($\rm
U',U'',U'''$).}
\label{Table3}
\begin{center}
\begin{tabular}{|l|rrrrr|rrrr|}\hline
&$\rm U_1$&$\rm U_2$&$\rm U_3$&$\rm U_4$&$\rm U_5$&$\rm U_A$
&$\rm U'$&$\rm U''$&$\rm U'''$\\
\hline
$T_1$&$-{1\over2}$&0&${1\over2}$&0&0&${5\over2}$&0&0&${1\over2}$\\
$T_2$&$-{1\over2}$&$-{1\over2}$&0&0&$-{1\over2}$&${5\over2}$&0&0&$-1$\\
$T_3$&$-{1\over2}$&0&${1\over2}$&0&0&${5\over2}$&0&0&${1\over2}$\\
\hline
\end{tabular}
\vspace{2cm}
\caption{The hidden SU(4) fields and their transformation properties
under $G_{\rm U(1)}$. $D_i$ represent sixplets ({\bf6}), whereas $\widetilde
F_i, \widetilde{\bar F}_i$ represent tetraplets (${\bf4},\bar{\bf4}$). Also
indicated are the charges under $\rm U_A$ and three orthogonal linear
combinations of interest ($\rm U',U'',U'''$).}
\label{Table4}
\medskip
\begin{tabular}{|l|rrrrr|rrrr|}\hline
&$\rm U_1$&$\rm U_2$&$\rm U_3$&$\rm U_4$&$\rm U_5$&$\rm U_A$
&$\rm U'$&$\rm U''$&$\rm U'''$\\
\hline
$D_1$&0&$-{1\over2}$&${1\over2}$&${1\over2}$&0&${3\over2}$&${1\over2}$&2&0\\
$D_2$&0&$-{1\over2}$&${1\over2}$&$-{1\over2}$&0&${3\over2}$&${1\over2}$&2&0\\
$D_3$&$-{1\over2}$&0&${1\over2}$&0&0&${5\over2}$&0&0&${1\over2}$\\
$D_4$&$-{1\over2}$&$-{1\over2}$&0&0&${3\over2}$&${3\over2}$&$-1$&2&0\\
$D_5$&0&$-{1\over2}$&${1\over2}$&0&0&${3\over2}$&${1\over2}$&2&0\\
$D_6$&0&${1\over2}$&$-{1\over2}$&0&0&$-{3\over2}$&$-{1\over2}$&$-2$&0\\
$D_7$&${1\over2}$&0&$-{1\over2}$&0&0&$-{5\over2}$&0&0&$-{1\over2}$\\
$\widetilde F_1$&$-{1\over4}$&${1\over4}$&$-{1\over4}$&0&$-{1\over2}$
&${1\over2}$&0&$-{9\over4}$&$-{1\over2}$\\
$\widetilde F_2$&${1\over4}$&${1\over4}$&$-{1\over4}$&0&${1\over2}$
&$-2$&$-{1\over2}$&${1\over4}$&${1\over2}$\\
$\widetilde F_3$&${1\over4}$&$-{1\over4}$&$-{1\over4}$&0&${1\over2}$
&$-{3\over2}$&$-{1\over2}$&${7\over4}$&0\\
$\widetilde F_4$&$-{1\over4}$&${3\over4}$&${1\over4}$&0&0
&${1\over2}$&0&$-{9\over4}$&1\\
$\widetilde F_5$&$-{1\over4}$&${1\over4}$&$-{1\over4}$&0&${1\over2}$
&$-{1\over2}$&$-1$&$-{1\over4}$&${1\over2}$\\
$\widetilde F_6$&$-{1\over4}$&${1\over4}$&$-{1\over4}$&0&$-{1\over2}$
&${1\over2}$&0&$-{9\over4}$&$-{1\over2}$\\
$\widetilde{\bar F}_1$
&$-{1\over4}$&${1\over4}$&${1\over4}$&${1\over2}$&$-{1\over2}$
&${3\over2}$&${1\over2}$&$-{7\over4}$&0\\
$\widetilde{\bar F}_2$
&$-{1\over4}$&${1\over4}$&${1\over4}$&$-{1\over2}$&$-{1\over2}$
&${3\over2}$&${1\over2}$&$-{7\over4}$&0\\
$\widetilde{\bar F}_3$
&${1\over4}$&$-{1\over4}$&${1\over4}$&0&$-{1\over2}$
&${1\over2}$&1&${1\over4}$&$-{1\over2}$\\
$\widetilde{\bar F}_4$
&$-{1\over4}$&${1\over4}$&${1\over4}$&0&$-{1\over2}$
&${3\over2}$&${1\over2}$&$-{7\over4}$&0\\
$\widetilde{\bar F}_5$
&$-{1\over4}$&$-{1\over4}$&${1\over4}$&0&$-{1\over2}$
&2&${1\over2}$&$-{1\over4}$&$-{1\over2}$\\
$\widetilde{\bar F}_6$
&$-{3\over4}$&${1\over4}$&$-{1\over4}$&0&0
&${3\over2}$&$-1$&$-{7\over4}$&0\\
\hline
\end{tabular}
\end{center}
\end{table}
\clearpage

\newpage
\clearpage

\newpage
\clearpage

\clearpage

\end{document}